\documentclass[aps,prd,notitlepage,superscriptaddress,reprint]{revtex4-1}
\usepackage{graphicx}
\usepackage[tbtags]{amsmath}
\usepackage{amssymb,amsthm} 
\usepackage{epstopdf,cancel,ulem}
\usepackage[colorlinks,bookmarks=false]{hyperref}

\newcommand{\be}{\begin{equation}}
\newcommand{\ee}[1]{\label{#1} \end{equation}}
\newcommand{\fracp}[2]{\frac{\partial #1}{\partial #2}}

\def\Qmu{Q^\mu}

\def\Qnu{Q^\nu}

\newcommand{\Xmu}[1]{x_{#1}^\mu}

\newcommand{\Xnu}[1]{x_{#1}^\nu}

\newcommand{\Xmp}[1]{x_{#1}^{\prime \mu}}

\newcommand{\Xnp}[1]{x_{#1}^{\prime \nu}}

\newcommand{\Xip}[1]{x_{#1}^{\prime i}}
\def\gmnp{g^\prime_{\mu\nu}}
\def\gop{{g^\prime}_{00}}

\def\gmn{g_{\mu\nu}}
\def\Gmn{g^{\mu\nu}}
\def\go{g_{00}}

\def\r{H_{rest}}

\def\td{\dot\tau}

\newcommand{\mean}[1]{\ensuremath{\left\langle #1 \right\rangle}}

\begin{document}

\author{Magdalena Zych}
\email{m.zych@uq.edu.au}
\affiliation{Centre for Engineered Quantum Systems, School of Mathematics and Physics, The University of Queensland, St Lucia, QLD 4072, Australia }
\author{{\L}ukasz Rudnicki}
\affiliation{Max-Planck-Institut f\"{u}r die Physik des Lichts, Staudtstra{\ss}e 2, 91058 Erlangen, Germany}
\affiliation{ Center for Theoretical Physics, Polish Academy of Sciences, Aleja Lotnik\'{o}w 32/46, 02-668 Warsaw, Poland}
\author{Igor Pikovski}
\affiliation{ITAMP, Harvard-Smithsonian Center for Astrophysics, Cambridge, MA 02138, USA}
\affiliation{Department of Physics, Harvard University, Cambridge, MA 02138, USA}
\affiliation{Department of Physics, Stevens Institute of Technology, Hoboken, NJ 07030, USA}

\title{Gravitational mass of composite systems}

\begin{abstract}
The equivalence principle in combination with the special relativistic equivalence between mass and energy, $E=mc^2$, is one of the cornerstones of general relativity.
However, for composite systems a long-standing result in general relativity asserts that the passive gravitational mass is not simply equal to the total energy.
This seeming anomaly is supported by all explicit derivations of the dynamics of bound systems, and is only avoided after time-averaging. Here we rectify this misconception and derive from first principles the correct gravitational mass
of a generic bound system in an external gravitational field. Our results clarify a lasting conundrum in general relativity and show how the weak and strong equivalence principles naturally manifest themselves for composite systems. The results
are crucial for describing new effects due to the quantization of the interaction between gravity and composite systems.
\end{abstract}

\maketitle
\section{Introduction} 
The equivalence principle postulates the exact equality between inertial and gravitational mass of any object,
regardless of composition.
This fundamental equality paved the way to a metric theory of gravity and is a vital pillar of general relativity \cite{Einstein:1916:FOGR, Will:2014}. It implies the universality of the gravitational interaction: all systems and forms of energy in a sufficiently small region of space are affected equally by gravity.
Many different experiments have confirmed this principle \cite{shapiro1999century,wagner2012torsion,williams2012lunar,Haensch:2004,Schlippert:2014,zhou2015test,touboul2017microscope,archibald2018universality}, with the most stringent bound on its violations currently being a few parts in $10^{-14}$ \cite{touboul2017microscope}. Recently, it was shown that the quantized gravitational interaction with composite systems yields novel effects and experiments \cite{Zych:2011, Zych:2012, Zych2016, margalit2015self, bushev2016single,pikovskiuniversal2015, ref:AB, PikovskiTime2017, korbicz2017information,zych2015quantum, zych2015PhD,rosi2017quantum,2016CQGra..33sLT01O, geiger2018proposal}, which rely on the coupling of gravity to the total energy of composite systems, as dictated by the equivalence principle.

Yet, general relativistic calculations for composite systems reveal an intricate dynamics which seems to be at odds with the equivalence principle and the mass-energy equivalence. The physical scenario is the coupling of a small composite system, such as a molecule, to the gravitational field of a much more massive object, like the Earth.
The passive gravitational mass of composite systems, i.e. the quantity coupling to the background gravitational potential of the post-Newtonian metric, is not given by its total energy. Rather, to first order in $c^{-2}$, the gravitational mass of an interacting $N$-particle system is derived to be \cite{eddington1938problem, nordtvedt1970gravitational, lightman1973restricted, nordtvedt1974equation, Nordtvedt:1994PostNewtonian} $M(G) = \sum_i^N (m_i + 3 m_i v_i^2/2c^2 - 2\sum_{j > i}^N k q_i q_j/r_{ij}c^2) $, where $k$ is the coupling between the particles, $q_i$ their charges for the specific interaction, $m_i$ their rest masses, $v_i$ their velocities, and $r_{ij}$ their relative distances. Gravity therefore seemingly does not simply couple to the rest, kinetic and potential energies, $R$, $T$ and $U$, respectively, but to $M(G) =(R+3T+2U)/c^2$.
This result was first noted by Eddington and Clark \cite{eddington1938problem} and has since been rederived in all explicit calculations, both for classical \cite{nordtvedt1970gravitational, lightman1973restricted, nordtvedt1974equation, Nordtvedt:1994PostNewtonian, Carlip:1999KE} and for quantized systems \cite{PhysRevD.23.2157, Lebed2013}.
The dynamics of composite systems does not take a single-particle form, i.e.\ the internal energies do not simply add to the gravitational mass in equal proportions, in apparent violation of the universality of the gravitational coupling.

For the active gravitational mass, the anomaly can be resolved by defining the mass in terms of its effects on test particles at spatial infinity, as in the ADM formalism~\cite{ADM:2008}. Such defined mass is equal to the total energy~\cite{Brill:1963Energy}. But the same method does not directly apply to the passive mass, for which the anomalous terms remain. A common resolution is to invoke time-averaging and the virial theorem~\cite{eddington1938problem, nordtvedt1970gravitational, lightman1973restricted, nordtvedt1974equation, Nordtvedt:1994PostNewtonian, Carlip:1999KE}, which yields $\mean{2T + U}=0$ and restores the expected coupling.
But the virial theorem does not imply fundamental validity, suggesting a violation of the mass-energy equivalence and of equality with the active mass, beyond the time-averaged dynamics. Even worse, the anomalous coupling would generically show up on the quantum level --
beyond the ensemble average. The 'virial terms' in the gravitational mass have lingered in the literature for decades and have led to the belief that the mass-energy equivalence may not exactly hold \cite{eddington1938problem, PhysRevD.23.2157}, as well as to specific experimental proposals to search for the violations \cite{Lebed2013}.

In this work we derive the gravitational coupling for composite systems
from first principles. We show, contrary to many previous results, that gravity only couples to the total internal energy of the bound system, as expected from the foundations of the theory.
We derive the passive gravitational mass for a generic composite system in curved space-time and show that this dynamics
takes a single-particle form~\footnote{Provided that tidal forces are negligible -- the usual assumption under which the equivalence principle is required to hold.}.
Crucial for isolating the correct gravitational coupling is to identify the physically correct internal energy, which removes the anomalous `virial terms'. Our resolution is in line with remarks made in ref.~\cite{Carlip:1999KE}, where it was noted that the virial theorem itself is a consequence of general covariance and the `virial terms'  must therefore be coordinate artifacts -- suggesting that a correct definition of the gravitational mass should be possible from first principles. Here we provide such a definition and clarify the physical meaning of the involved coordinates.
We demonstrate our general framework in explicit examples that show how the correct gravitational mass emerges for electromagnetically and gravitationally bound systems.

\section{Gravitational coupling to bound systems}
\subsection{Lagrangian in two sets of coordinates}
In our analysis the metric tensor $\gmn$, $\mu, \nu=0,...,3$  is fixed and has signature $(-+++)$ and {describes} a static symmetric space-time, with $g_{0i} = g_{i0} = 0$ and $g_{ij} = g_{ji}$ for $i,j = 1,2,3$.
For a single particle with mass $m$, on a world line $x=x^\mu(s)$, where $s$ is an arbitrary parameter, the Lagrangian is \cite{WeinbergGR} $L=-mc^2\frac{d\tau}{ds}$, where ${d\tau}=c^{-1}\sqrt{-\gmn(x)d\Xmu{}d\Xnu{}}$ is an infinitesimal proper time element along the world line.
We consider a closed system of $N$ interacting particles that can be described by a Lagrangian $L_N$. For example, for electromagnetic interactions the Lagrangian reads  \cite{WeinbergGR}:
\be
L_N =\sum_n \left( -m_nc^2\frac{d\tau_n}{ds}+e_nA_\mu(x_n)\frac{d\Xmu{n}(s)}{ds} \right),
\ee{LN}
where $m_n$,  $e_n$ and $\Xmu{n}(s)$ for  $n = 1,..,N$ describe the mass, charge and world line of the $n^{th}$ particle, respectively, and $A_\mu(x_n)$ is the electromagnetic four-potential at $x_n$, produced by all particles. This Lagrangian describes interacting particles without emission of radiation, i.e.~to order $c^{-2}$ such that the field degrees of freedom (DOF) and retardation effects can be neglected \cite{Landau:1975}. We can choose $x^0_n \equiv s$ for all $n$ and identify $s\equiv ct$, so that $t$ is the coordinate time~\footnote{The parameter $t$ is operationally  defined as the time measured by a clock at rest in the origin of the reference frame with respect to which the spatial coordinates of the particles are defined.};
we will denote the derivative with respect to $t$ as $\dot a:=\frac{da}{dt}$.

\begin{figure} [t]
\centering
\includegraphics[width=1.01\columnwidth]{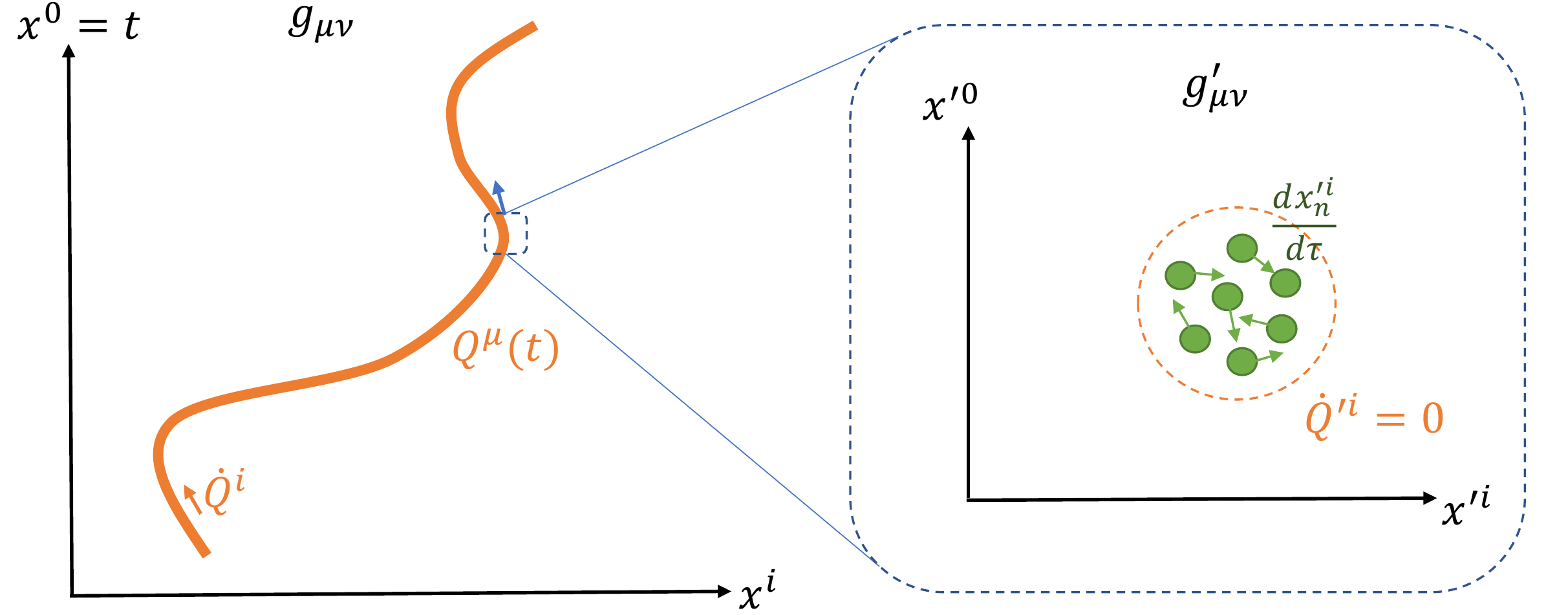}
\caption{A bound N-particle system on a world-line $\Qmu(t)$. The primed coordinates describe the frame in which the CM is at rest, under the conditions \eqref{metricapprox}. This frame defines the physical internal properties of the system and the correct gravitational mass. }
\label{Fig:1}
	\end{figure}

Let us pick an arbitrary world line $\Qmu(t)$ and define new coordinates $Q^{\prime \mu}= \fracp{\Xmp{}}{\Xnu{}}Q^\nu$ relative to $\Qmu$ in the sense that $\dot Q^{\prime i}=0$, and such that $Q^{\prime0}$ is the proper time along this world line: $Q^{\prime0}=
c^{-1}\int dt\big({-\gmn(Q)\dot\Qmu\dot\Qnu}\big)^{1/2}\equiv\tau$, see Fig.~\ref{Fig:1}.
Eq.~\eqref{LN} in terms of $\tau$ and $t$
reads
\be
L_N \!=\!\sum_n \bigg( \!-m_n c^2 \sqrt{-\gmnp\frac{d\Xmp{n}}{d\tau}\frac{d\Xnp{n}}{d\tau}}+  e_nA^{\prime}_{\mu}\frac{d\Xmp{n}}{d\tau}\bigg)\td
\ee{LN2}
Eq.~\eqref{LN2} is exactly the same as eq.~\eqref{LN}, but  uses two sets of coordinates: the original ones for describing the arbitrary world line $Q^\mu$ through $c\td=\big({-\gmn(Q)\dot\Qmu\dot\Qnu}\big)^{1/2}$, and the primed ones for describing the system relative to $Q^\mu$.
The Lagrangian has now the product form $L_N=L^{\prime}\!\cdot\! \td$, in direct analogy to the relativistic single particle Lagrangian $L_1=-mc^2\td$.

The use of two different sets of coordinates within the same description is key to correctly predict the physical effects in bound systems, as will be shown below. We note that such a procedure has also been used in other contexts within general relativity to describe physical effects correctly  \cite{damour1987problem,Brumberg1989,PhysRevD.43.3273,PhysRevD.45.1017,PhysRevD.49.618}. In particular,  a similar observation has been made in the context of celestial mechanics by T.~Damour~\cite{damour1987problem}, who pointed out that certain deformations of a body from  spherical symmetry reported by other authors are artifacts  of using an external coordinate system to define local properties. Such a treatment implicitly introduces Lorentz and gravitational redshift factors which deform the system, nonetheless, the deformations are not intrinsic to the body and are removed by the correct choice of local coordinates.  This is analogous to the topic studied in the present work, and we thus proceed to show explicitly  how the apparent “deformations” of the gravitational mass of a composite system allegedly violating the equivalence principle, reported in previous studies,  are removed when the gravitational mass is described in the correct, local coordinates.

\subsection{Conditions for defining a centre of mass}
We now seek to identify $Q^\mu$ with the world line of the composite system -- its centre of mass (CM) -- and the primed coordinates with the centre of momentum frame, in which the CM is at rest. However, the canonical momentum conjugate to $x_n$ is $\fracp{L_N}{\dot x^i_n}$
and generally there is no unique way of defining the total linear momentum since the individual particle momenta belong to different tangent spaces \cite{Dixon:1964}. A total momentum can nevertheless be consistently defined when the metric is
approximately constant in the region occupied by all $N$ constituents
\be
\forall_{n,m}\,  \gmn(x_n)\approx\gmn(x_m), \quad \forall_{n,m} \, x_n\approx x_m,
\ee{metricapprox}
The first condition means that the space-time in the region occupied by the system is approximately flat and a single coordinate system can be introduced in which the metric is locally the Minkowski metric $\eta_{\mu\nu}$. For well-behaved metrics (e.g.~if the metric components are Lipschitz functions) this condition is satisfied if the individual world lines are sufficiently close, the second condition in eq.~\eqref{metricapprox}.
Eqs.~\eqref{metricapprox} imply that tidal effects between the particles can be neglected, which allows the construction of
a generally covariant notion of the CM for a generic extended system, as explicitly shown in~\cite{Dixon:1964, beiglbock1967}.

With the individual linear momenta denoted by $P_{ni}(x_n)$, under the assumption \eqref{metricapprox} the total momentum is $P_i=\sum_n P_{ni}(x_n)$. The centre of momentum  (primed) frame is defined by $P^{\prime}_i=\sum_n P^{\prime}_{ni}(x_n)=0$, where $P^{\prime}_{ni}=\fracp{\Xmu{}}{\Xip{}}P_{n\mu}$.
If $x_n^\mu(t)$ are world lines of the individual constituents of the system, the error made in describing the $N$-particle system as a single composite particle following a world line $x^\mu(t)$ can be  quantified by the difference between the sum of the contravariant momenta: one where the metric used to raise the indices is evaluated at different points and one where
the metric is evaluated in a single point 
\be
 P^\mu - \Gmn(x)P_\nu=\sum_n(\Gmn(x_n)-\Gmn(x))P_{n\nu}.
\ee{error}
The approximations~\eqref{metricapprox} depend on the variation of the metric across the region occupied by the constituent particles as compared to the energy-momentum of the system. Consider a region
$\mathcal U := \bigcup_t \mathcal U_t$, with $\mathcal U_t$ such that $\forall_n x_n(t)\in \mathcal U_t$. Assuming the variation of the metric in $\mathcal U$ is bounded can be expressed as
\be
\exists_{K>0} \forall_{\mu\nu, n, m}| \Gmn(x_n)-\Gmn(x_m)| <K \, .
\ee{bound_on_g}
For example, this is satisfied by the Schwarzschild metric in isotropic coordinates, whose components are Lipschitz functions in any compact space-time region with no singularity.
If the four-momenta of the particles in the considered region are bounded, we can define
\be
\tilde P := \mathrm{max}\{ |P_{n\mu}(x_n)| : n \in \{1,...,N\}, x_n \in \mathcal U, \mu = 0,...,3\}.
\ee{bound_on_P}
Using eqs.~\eqref{bound_on_g} and \eqref{bound_on_P},  the magnitude of the error, eq.~\eqref{error}, satisfies
\be
|P^\mu - \Gmn P_\nu | =|\sum_{n,\nu}(\Gmn(x_n)-\Gmn(x))P_{n\nu}|< 4 N K \tilde P,
\ee{conditions}
for all $\mu$.
We note that eq.\ \eqref{conditions} means that if the energy of the system is finite, and given a finite measurement precision, for any composite system of relativistic particles (on a well-behaved metric) there exist a bound on the volume occupied by the system, such that the error made by using the approximation in eq.~(3) in the main text is below the measurement precision, as long as the system's size is smaller than this bound.

Under the approximations \eqref{metricapprox}, we can choose $\Qmu(t)$ to be specifically the world line of the CM of the N-particle system, and the primed coordinates to be the centre of momentum frame.
Eq.~\eqref{LN2} can now be expressed in terms of the rest frame Lagrangian of the N-particle system, i.e.~the Lagrangian of the internal DOF in the CM rest frame. {This can be an arbitrary Lagrangian $L_{rest}$, for the specific example \eqref{LN} it reads}:
\be
L_{rest}=\sum_n \bigg(-m_nc^2\sqrt{-\gmnp\frac{d\Xmp{n}}{d\tau}\frac{d\Xnp{n}}{d\tau}}+e_nA^{\prime}_{\mu}\frac{d\Xmp{n}}{d\tau}\bigg)
\ee{Lrest_comp}
with $\gmnp=\eta_{\mu\nu}$. The total Lagrangian in the presence of gravity is thus simply 

\be
L_N\approx L_{rest}\td,
\ee{complag}
where $\tau$ is the proper time along the CM world line.

\subsection{Hamiltonian for bound systems on a background space-time}
Lagrangian \eqref{complag} has a single-particle form, with $-mc^2$ generalised to $L_{rest}$, which suggests that the total mass of the 
system is defined dynamically and is given by the total internal energy. This is  explicitly seen in the Hamiltonian picture. The Legendre transform of eq.~\eqref{complag} yields $H_{N}=P_i\dot Q^i+\sum_n p^{\prime}_{in}\dot x^{\prime i}_n -L_{N}$, where $P_i$ is the canonical momentum associated with the CM coordinate $Q^i$,  while $p_n^{\prime}$ are the internal momenta, canonically conjugate to the internal DOFs in the system's rest frame:
\be
P_i:=\fracp{L_{N}}{\dot Q^i} , \quad p_{in}^{\prime}:= \fracp{L_{rest}}{\frac{dx^{\prime i}_n}{d\tau}}.
\ee{PN}
The rest frame Hamiltonian is by definition
\be
\r=\sum_n \fracp{L_{rest}}{\frac{dx^{\prime}_n}{d\tau}}\frac{dx^{\prime}_n}{d\tau}- L_{rest}.
\ee{Hrest}
The explicit expression for the external momentum is found by using eq.~\eqref{complag} in eq.~\eqref{PN}: $P_i=\fracp{L_{rest}}{\dot Q^i}\td+L_{rest}\frac{d\td}{d\dot Q^i}$.  The simple equality $\frac{dx^i}{d\tau}=\frac{\dot x^i}{\td}$ further yields $\frac{L_{rest}}{d\dot Q^i}=-\sum_n\fracp{L_{rest}}{\frac{dx^i_n}{d\tau}}\dot x^i_n\frac{1}{\td^2}\frac{d\td}{d\dot Q^i}$ and thus
 \be
P_i=\bigg(\sum_n-\fracp{L_{rest}}{\frac{x^\prime_n}{d\tau}} \frac{dx^\prime_n}{d\tau}+L_{rest}\bigg)\frac{d\td}{d\dot Q^i}=\r\frac{\dot Q_i}{c^2\td},
\ee{cm_mom}
where  $c\td=\sqrt{-\gmn(Q)\dot\Qmu\dot\Qnu}$ and $\r$ is given by eq.~\eqref{Hrest}.
Substituting the above into the Legendre transform for the total Hamiltonian and using the definition of $p_n$ gives $H_{N}=\r\frac{\dot Q^i\dot Q_i}{c^2\td}+\r\td$. Using $ Q^0:=ct$ yields
\be
 H_{N}=-\r\frac{\go}{\td}.
\ee{HN_app}
From eq.~\eqref{cm_mom} we next find
\be
c^2P_iP^i=\r^2\frac{\dot Q^i\dot Q_i}{c^2\td^2}=\r^2(-1+\frac{-\go}{\td^2}),
\ee{cm_square}
which upon substitution into eq.\ \eqref{HN_app} yields eq.~(8) in the total N-particle Hamiltonian for the system:
\be
H_{N}=\sqrt{-\go(c^2P_iP^i+\r^2)}.
\ee{HN2}
The above result entails that a many-particle system following a narrow world-tube (satisfying eq.~\eqref{metricapprox}) is effectively described as a composite particle whose total mass is $\r/c^2$, where $\r$ is the rest frame energy of the system.
This is in explicit agreement with the equivalence principle and in particular confirms that the (passive) gravitational mass of a composite system is equal to its total internal energy in appropriate units.  For $L_{rest}$ in eq.\ \eqref{Lrest_comp},  $\r= \sum_n (c\sqrt{-\gop[(p_{ni}^{\prime}-e_nA^\prime_i)(p_n^{\prime i}-e_n A^{\prime i})+m_n^2c^2]}- e_n A^\prime_0 )$, where
$p_n^{\prime i}=m_n\frac{ dx^{\prime}_{ni}}{d\tau_n}+e_nA^\prime_i$.

\section{Specific examples}
We now apply our result to several scenarios previously discussed in the literature.
We show that the choice of the correct (rest frame) coordinates for the internal DOFs fully resolves any apparent tension between the equivalence principle and the general relativistic description of composite systems.

\subsection{Example i: Hydrogen-like system}
A special case of eq.~\eqref{LN} was considered in refs.~\cite{nordtvedt1970gravitational, lightman1973restricted, nordtvedt1974equation, Nordtvedt:1994PostNewtonian, PhysRevD.23.2157, Lebed2013}. The composite system here comprises two charges interacting via the Coulomb potential on a post-Newtonian metric
\be
g_{00}=-\big(1+2\frac{\phi(x)}{c^2}\big), \quad g_{ij} =\delta_{ij}\big(1-2\frac{\phi(x)}{c^2}\big),
\ee{pN_metric}
where $\phi(x)$ is the external gravitational potential.
On a flat metric, the non-relativistic Lagrangian for this system is $L=\sum_{i=1}^2 \left( -m_i c^2 + m_i\dot{\vec{x}}_i^2/2 \right) - k e_1e_2/|\vec{x}_1-\vec{x}_2|$, with $\vec x\equiv (x^1, x^2, x^3)$ and the Coulomb's constant $k$. It applies to slowly moving particles as it ignores special-relativistic kinetic terms and magnetic interactions between charges in relative motion. Therefore, one can define the usual CM and relative coordinates, respectively: $\vec R:=\sum m_i\vec{x}_i/M$, $\vec{r}:= \vec{x}_1-\vec{x}_2$ and $\vec v:=\dot{\vec r}$, with $M:=\sum_i m_i$ and $\mu:= m_1 m_2/M$. The Lagrangian of the system in the CM rest frame is
 \be
L_{rest}=-Mc^2 +\frac{\mu{\vec{v}}^{\prime 2}}{2} -{k}\frac{e_1e_2}{r^{\prime}},
\ee{L2_fin}
Eqs.~\eqref{complag}--\eqref{Hrest} yield $H_2=\vec p^{\prime} \vec v^{\prime}\td-L_2\equiv\left(\vec p^{\prime}\vec v^{\prime}-L_{rest}\right)\td$ where by definition $\left(\vec p^{\prime}\vec v^{\prime}-L_{rest}\right)\equiv\r$ and where $\td=1+\frac{\phi(x)}{c^2}$. Thus the Hamiltonian for the system subject to gravity on the space-time metric~\eqref{pN_metric} is
\be
H_2=\big[Mc^2 +\frac{\vec p^{\prime 2}}{2\mu} +k\frac{e_1e_2}{r^{\prime}}\big]\big(1+\frac{\phi}{c^2}\big).
\ee{H2_fin}
The gravitational mass of the system, i.e.~the quantity coupling to $\phi$, is the total energy in the CM rest frame $Mc^2 +T_{rest} +U_{rest}$, with $T_{rest}=\frac{{\vec p^{\prime 2}}}{2\mu}$ and $U_{rest}=k\frac{e_1e_2}{r^{\prime}}$, in explicit agreement with the equivalence principle.

This result seems to be at odds with previous studies \cite{nordtvedt1970gravitational, lightman1973restricted, nordtvedt1974equation, Nordtvedt:1994PostNewtonian, PhysRevD.23.2157, Lebed2013}, where the coupling (without time-averaging) takes a different form. However, we now show that the dynamics is exactly the same and that the anomalous couplings found previously are coordinate artifacts. To clarify this, we repeat the derivation using the Lagrangian expressed only in terms of the external coordinates that define the metric \eqref{pN_metric}, as in previous works:
\be
\begin{split}
L_2&=\sum_{i=1,2} \left[-m_ic^2\big(1+\frac{\phi(x_i)}{c^2}\big)+\frac{m_i\dot{\vec{x}}_i^2}{2}\big(1-3\frac{\phi(x_i)}{c^2}\big)\right]\\
- \frac{k}{2} & \frac{e_1e_2}{|\vec{x}_1-\vec{x}_2|}\big(1+2\frac{\phi(x_1)}{c^2}\big) -\frac{k}{2} \frac{e_1e_2}{|\vec{x}_1-\vec{x}_2|}\big(1+2\frac{\phi(x_2)}{c^2}\big).
\end{split}
\ee{L2}
Eq.~\eqref{L2} in terms of the CM and relative coordinates is
 \be
L_2\!=\!-Mc^2\big(1+\frac{\phi}{c^2}\big) +\frac{\mu{\vec{v}}^2}{2}\big(1-3\frac{\phi}{c^2}\big)
 -k\frac{e_1e_2}{r}\big(1+2\frac{\phi}{c^2}\big),
\ee{L2_2}
where we assumed for clarity that the CM is stationary, $\dot{\vec R}\approx 0$, and used eqs~\eqref{metricapprox} to set $\phi(x_i)\approx \phi(R)\equiv\phi$. The apparent challenge to the equivalence principle arises from Lagrangian  \eqref{L2_2} and the corresponding Hamiltonian. The canonical momentum is $\vec p=\mu \vec v\big(1-3\frac{\phi}{c^2}\big)$
and the Legendre transform of eq.~\eqref{L2_2} yields
\be
H_2=Mc^2\big(1+\frac{\phi}{c^2}\big) + \frac{{\vec p}^2}{2\mu}\big(1+3\frac{\phi}{c^2}\big)
+k\frac{e_1e_2}{r}\big(1+2\frac{\phi}{c^2}\big),
\ee{H2}
which features the anomalous coupling of the gravitational potential to $3T+2U$ with $T=\frac{{\vec p}^2}{2\mu}$ and $U=k\frac{e_1e_2}{r}$.
However, both $T$ and $U$ are here expressed  in the original coordinates which can be interpreted as local coordinates of a \textit{distant} observer. $T$ and $U$ thus include the redshift factors that depend on the choice of this distant observer and do not describe the 
\textit{local}, physical quantities in the rest frame of the system. Therefore they cannot be interpreted as the internal kinetic and potential energies of the bound system.

We now show how to amend eq.~\eqref{H2} and find the physically correct internal energies. The local distance $d\vec{x}^{\prime}$ and the coordinate distance $d\vec{x}$ on the metric \eqref{pN_metric} satisfy $d{x}^{\prime i}\approx\big(1-\frac{\phi}{c^2}\big)d{x}^i$; whereas the local (proper) time and the coordinate time $t$ satisfy $d\tau=dt^{\prime}\approx\big(1+\frac{\phi}{c^2}\big)dt$. This yields $\vec v=\frac{d\vec x}{dt}=\vec v^{\prime} \big(1+2\frac{\phi}{c^2}\big)$, where $\vec v^{\prime}:=\frac{d\vec{x}^{\prime}}{d\tau}$ is the velocity of the relative DOF in the local rest frame of the CM.
The momentum thus satisfies  $\vec p = \vec{p}^{\prime} \big(1-\frac{\phi}{c^2}\big)$, where $\vec p^{\prime}=\fracp{L_{rest}}{ \vec v^{\prime}}=\mu\vec v^{\prime}$. The internal kinetic energy is $T_{rest}=\frac{{\vec p^{\prime 2}}}{2\mu}$ and we find
\be
 T\big(1+3\frac{\phi}{c^2}\big) = T_{rest} \big(1+\frac{\phi}{c^2}\big).
\ee{TT02}
Denoting by $r^{\prime}$ the distance between the two charges in the CM rest frame yields $r^{\prime}=\big(1-\frac{\phi}{c^2}\big)r$, and thus the rest frame potential energy $U_{rest}=k\frac{e_1e_2}{r^{\prime}}$ satisfies
 \be
U\big(1+2\frac{\phi}{c^2}\big) = U_{rest} \big(1+\frac{\phi}{c^2}\big).
\ee{UU0}
Using eqs~\eqref{UU0} and \eqref{TT02}, Hamiltonian~\eqref{H2} reads
\be
H_2=\big[Mc^2 +T_{rest} +U_{rest}\big]\big(1+\frac{\phi}{c^2}\big),
\ee{H2_fin2}
in agreement with our derivation, eq. \eqref{H2_fin}. The correct expression for the gravitational mass is now apparent because the CM rest frame coordinates are used to describe the internal DOFs, while external coordinates are used to capture the coupling of the CM to gravity.

\subsection{Example ii: Gravitationally bound systems and the strong equivalence principle}
We now consider a system bound only through gravity, in the presence of a background metric produced by a much larger mass. According to the strong equivalence principle, such a system should couple to gravity in the same way as any other composite system. In the Newtonian approximation,
the Lagrangian \eqref{L2_fin} describes a gravitationally bound system with the replacement $-k e_1 e_2 \rightarrow G m_1 m_2$ for the interaction. This yields the Hamiltonian
\be
H_2^{G} = \big[Mc^2 +\frac{\vec{p}^{\prime 2}}{2\mu} - G\frac{M \mu}{r^{\prime}}\big]\big(1+\frac{\phi}{c^2}\big),
\ee{eq:H2GravNewton}
where $M=m_1+m_2$, $\mu =m_1 m_2/M $, as before. Thus a bound system has an effective gravitational mass that includes the gravitational binding energy, an explicit confirmation of the strong equivalence principle. Note that this differs from the result obtained by Eddington and Clark \cite{eddington1938problem}, which has the additional anomalous `virial terms', an artefact of using redshifted coordinates to describe the internal energy as discussed above.

Going beyond the Newtonian limit, in the weak-field approximation and for slowly moving particles one can extend the analysis to a bound system fully described by general relativity. Such a system was first considered by Einstein, Infeld and Hoffmann \cite{einstein1938gravitational} and by Eddington and Clark \cite{eddington1938problem}. A Lagrangian can be defined if emission of radiation is neglected, i.e. to orders below $c^{-5/2}$.
The previous studies considered the $N$-particle system on a flat background space-time, i.e. each particle $i$ producing a field $g_{\mu \nu}^{(i)} = \eta_{\mu \nu} + h_{\mu \nu}^{(i)}$. Here we are interested in the coupling of the entire system to the metric produced by a large external mass, thus the particle interactions are to be described on top of this external metric $g_{\mu \nu} \neq \eta_{\mu \nu}$. However, the approximations \eqref{metricapprox} ensure that we can choose a primed coordinate system in which the background metric becomes flat, $g'_{\mu \nu} = \eta_{\mu \nu}$, over the extension of the entire N-particle system. We can thus apply previous results in the CM rest frame~\footnote{To this order of approximation, the coordinates for the relativistic center of inertia are well-defined}, and include  the coupling to the external field through a coordinate transformation. The 2-particle Hamiltonian for a gravitationally bound system to order $c^{-2}$, and in the presence of a background metric becomes
\be
\begin{split}
& H_2^{GR} \!   = \! \left[ \frac{p^{\prime 2}}{2 \mu} \! \left(\! 1 -  p^{\prime 2}\frac{M \! - \! 3\mu}{4 c^2 \mu^2} \! \right)  -  G \frac{\mu M}{r^{\prime}} \left(1 \! - \! \frac{G M}{2c^2r^{\prime}} \! \right)  \right. \\
 & \, \, \left.  -  \frac{G}{2 c^2 r^{\prime}} \left( p^{\prime 2} \frac{3 M  +  \mu}{\mu}  +  \frac{\left(\vec{p}^{\prime}  \cdot  \vec r^{\prime} \right)^2}{r^{\prime 2}} \right)  +Mc^2  \right]  \frac{d\tau}{dt}
\end{split}
\ee{eq:H2Grav}
The first factor is the total energy of the two-body system in its CM rest frame, including relativistic corrections, while the factor $d\tau/dt$ captures the coupling to the external gravitational field. This has again the single particle form as required by the equivalence principle. Hamiltonian \eqref{eq:H2Grav} reduces to \eqref{eq:H2GravNewton} in the non-relativistic limit of the CM system and to lowest order in the coupling to the external gravitational potential, $d\tau/dt \approx (1+\phi/c^2)$.

\subsection{Example iii: Box of photons}
Another system, studied in ref.~\cite{Carlip:1999KE}, is a slowly moving `box of photons', where the internal energy is the kinetic energy of the box, $T$, and the energy of light, $U^{light}$. Variation of the matter action on the metric \eqref{pN_metric} yields the total energy \cite{Carlip:1999KE}
\be
 E=T\big(1+3\frac{\phi}{c^2}\big)+U^{light}\big(1+2\frac{\phi}{c^2}\big),
\ee{TUlight}
which again features the anomalous coupling terms.
Due to eq.~\eqref{TT02}, to find the physical coupling to gravity we only need to show that in the local rest frame of the box eq.~\eqref{UU0} holds for $U^{light}$. In generic coordinates $U^{light}=\int d^3x \sqrt{-g}\,T^{00}$, where $T^{00}$ is the relevant component of the energy-momentum tensor of the electromagnetic field and $g=\mathrm{Det}g_{\mu\nu}$.
In the rest frame of the box $U^{light}_{rest}=\int d^3x^{\prime} \sqrt{-\eta}\, T^{00}_{rest}$.  To lowest post-Newtonian order $T^{00}_{rest}=(1+2\frac{\phi}{c^2})T^{00}$, $d^3x^{\prime}=(1-3\frac{\phi}{c^2})d^3x$, and $\sqrt{-g}=(1-2\frac{\phi}{c^2})$. Thus $U^{light}=(1-\frac{\phi}{c^2})U^{light}_{rest}$, as required. Combined with
eq.~\eqref{TT02},
eq.~\eqref{TUlight} becomes
\be
E=\left(T_{rest} + U^{light}_{rest}\right)  \big(1+\frac{\phi}{c^2}\big),
\ee{TUlight2}
which explicitly satisfies the equivalence principle. Indeed, it was pointed out in ref.~\cite{Carlip:1999KE} that the additional terms in eq. \eqref{TUlight} are gauge artifacts. Here we have explicitly shown that correctly defining internal energies yields the true and unique gravitational mass and exposes the validity of the equivalence principle.

\section{Conclusions} This article shows how the correct gravitational mass emerges and how the equivalence principle manifests itself for composite systems. While no issue would be expected on an abstract level of the theory, it is surprising that all detailed calculations to date predict a coupling inconsistent with the equivalence principle, with experimentally measurable consequences. The correct physical picture is akin to the case for an elementary particle, for which by definition the mass is the total rest-frame energy. The same holds for a composite system: the mass is the total energy in its CM rest frame. To describe a composite system subject to gravity and isolate the physically relevant gravitational coupling, two different sets of coordinates are therefore invoked concurrently: arbitrary, external coordinates to describe the CM, and the CM rest-frame coordinates to describe the internal DOFs. This settles a long-standing issue with the passive gravitational mass of composite systems, which has been thought to include additional terms that only vanish on average and that violate the equivalence principle. Our results also demonstrate from first principles that the passive mass, being equal to the total energy content, is in accordance with the active (ADM) mass, as expected from the foundations of general relativity.

The study of quantum optical systems on curved space-time has recently become an active field of research, both in theory \cite{Zych:2011, Zych:2012, Zych2016, pikovskiuniversal2015, ref:AB, PikovskiTime2017, korbicz2017information, scully2018quantum} and experiment \cite{margalit2015self,rosi2017quantum,schlippert2014quantum}. Isolating the correct gravitational coupling for composite systems is  thus crucial for accurate predictions in upcoming quantum experiments which are starting to probe the interplay between quantum theory and general relativity. While all current classical tests are insensitive to the previously predicted anomalous couplings, the quantization of both internal and external DOFs reveals additional phenomena which depend on the correct form of the interaction \cite{Zych:2011,pikovskiuniversal2015}. Results of this work are thus central for upcoming probes of new effects, which include the time dilation induced entanglement between internal and spatial degrees of freedom \cite{Zych:2011, Zych:2012, Zych2016, margalit2015self, bushev2016single}, decoherence universally affecting composite quantum systems subject to time dilation \cite{pikovskiuniversal2015, ref:AB, PikovskiTime2017, korbicz2017information}, friction of relativistic decaying atoms \cite{sonnleitner2017will, sonnleitner2018mass} and quantum tests of the equivalence principle for composite systems~\cite{zych2015quantum, rosi2017quantum, 2016CQGra..33sLT01O, geiger2018proposal}.

We thank S.~Braunstein, \v{C}.~Brukner, F.~Costa and S.~Deser for helpful discussions and comments. M.Z.~acknowledges support through an ARC DECRA grant DE180101443, and ARC Centre EQuS CE170100009. {\L}.R. acknowledges financial support from Grant No.~2014/13/D/ST2/01886 of the National Science Center, Poland.
I.P.\ thanks the WACQT at Chalmers University for the kind hospitality and acknowledges support of the NSF through a grant to ITAMP and the Branco Weiss Fellowship -- Society in Science, administered by the ETH Z\"{u}rich.  This publication was made possible through the support of a grant from the John Templeton Foundation. The opinions expressed in this publication are those of the authors and do not necessarily reflect the views of the John Templeton Foundation. M.Z.~acknowledges the traditional owners of the land on which the University of Queensland is situated, the Turrbal and Jagera people.


\begin{thebibliography}{51}%
\makeatletter
\providecommand \@ifxundefined [1]{%
 \@ifx{#1\undefined}
}%
\providecommand \@ifnum [1]{%
 \ifnum #1\expandafter \@firstoftwo
 \else \expandafter \@secondoftwo
 \fi
}%
\providecommand \@ifx [1]{%
 \ifx #1\expandafter \@firstoftwo
 \else \expandafter \@secondoftwo
 \fi
}%
\providecommand \natexlab [1]{#1}%
\providecommand \enquote  [1]{``#1''}%
\providecommand \bibnamefont  [1]{#1}%
\providecommand \bibfnamefont [1]{#1}%
\providecommand \citenamefont [1]{#1}%
\providecommand \href@noop [0]{\@secondoftwo}%
\providecommand \href [0]{\begingroup \@sanitize@url \@href}%
\providecommand \@href[1]{\@@startlink{#1}\@@href}%
\providecommand \@@href[1]{\endgroup#1\@@endlink}%
\providecommand \@sanitize@url [0]{\catcode `\\12\catcode `\$12\catcode
  `\&12\catcode `\#12\catcode `\^12\catcode `\_12\catcode `\%12\relax}%
\providecommand \@@startlink[1]{}%
\providecommand \@@endlink[0]{}%
\providecommand \url  [0]{\begingroup\@sanitize@url \@url }%
\providecommand \@url [1]{\endgroup\@href {#1}{\urlprefix }}%
\providecommand \urlprefix  [0]{URL }%
\providecommand \Eprint [0]{\href }%
\providecommand \doibase [0]{http://dx.doi.org/}%
\providecommand \selectlanguage [0]{\@gobble}%
\providecommand \bibinfo  [0]{\@secondoftwo}%
\providecommand \bibfield  [0]{\@secondoftwo}%
\providecommand \translation [1]{[#1]}%
\providecommand \BibitemOpen [0]{}%
\providecommand \bibitemStop [0]{}%
\providecommand \bibitemNoStop [0]{.\EOS\space}%
\providecommand \EOS [0]{\spacefactor3000\relax}%
\providecommand \BibitemShut  [1]{\csname bibitem#1\endcsname}%
\let\auto@bib@innerbib\@empty
\bibitem [{\citenamefont {Einstein}(1916)}]{Einstein:1916:FOGR}%
  \BibitemOpen
  \bibfield  {author} {\bibinfo {author} {\bibfnamefont {A.}~\bibnamefont
  {Einstein}},\ }\href@noop {} {\bibfield  {journal} {\bibinfo  {journal}
  {Annalen der Physik}\ }\textbf {\bibinfo {volume} {40}},\ \bibinfo {pages}
  {284} (\bibinfo {year} {1916})}\BibitemShut {NoStop}%
\bibitem [{\citenamefont {Will}(2014)}]{Will:2014}%
  \BibitemOpen
  \bibfield  {author} {\bibinfo {author} {\bibfnamefont {C.~M.}\ \bibnamefont
  {Will}},\ }\href {\doibase 10.12942/lrr-2014-4} {\bibfield  {journal}
  {\bibinfo  {journal} {Living Reviews in Relativity}\ }\textbf {\bibinfo
  {volume} {17}},\ \bibinfo {pages} {4} (\bibinfo {year} {2014})},\ \Eprint
  {http://arxiv.org/abs/1403.7377} {arXiv:1403.7377 [gr-qc]} \BibitemShut
  {NoStop}%
\bibitem [{\citenamefont {Shapiro}(1999)}]{shapiro1999century}%
  \BibitemOpen
  \bibfield  {author} {\bibinfo {author} {\bibfnamefont {I.~I.}\ \bibnamefont
  {Shapiro}},\ }\href {\doibase 10.1103/RevModPhys.71.S41} {\bibfield
  {journal} {\bibinfo  {journal} {Reviews of Modern Physics}\ }\textbf
  {\bibinfo {volume} {71}},\ \bibinfo {pages} {S41} (\bibinfo {year}
  {1999})}\BibitemShut {NoStop}%
\bibitem [{\citenamefont {Wagner}\ \emph {et~al.}(2012)\citenamefont {Wagner},
  \citenamefont {Schlamminger}, \citenamefont {Gundlach},\ and\ \citenamefont
  {Adelberger}}]{wagner2012torsion}%
  \BibitemOpen
  \bibfield  {author} {\bibinfo {author} {\bibfnamefont {T.~A.}\ \bibnamefont
  {Wagner}}, \bibinfo {author} {\bibfnamefont {S.}~\bibnamefont
  {Schlamminger}}, \bibinfo {author} {\bibfnamefont {J.}~\bibnamefont
  {Gundlach}}, \ and\ \bibinfo {author} {\bibfnamefont {E.~G.}\ \bibnamefont
  {Adelberger}},\ }\href {\doibase 10.1088/0264-9381/29/18/184002} {\bibfield
  {journal} {\bibinfo  {journal} {Classical and Quantum Gravity}\ }\textbf
  {\bibinfo {volume} {29}},\ \bibinfo {pages} {184002} (\bibinfo {year}
  {2012})}\BibitemShut {NoStop}%
\bibitem [{\citenamefont {Williams}\ \emph {et~al.}(2012)\citenamefont
  {Williams}, \citenamefont {Turyshev},\ and\ \citenamefont
  {Boggs}}]{williams2012lunar}%
  \BibitemOpen
  \bibfield  {author} {\bibinfo {author} {\bibfnamefont {J.~G.}\ \bibnamefont
  {Williams}}, \bibinfo {author} {\bibfnamefont {S.~G.}\ \bibnamefont
  {Turyshev}}, \ and\ \bibinfo {author} {\bibfnamefont {D.~H.}\ \bibnamefont
  {Boggs}},\ }\href {\doibase 10.1088/0264-9381/29/18/184004} {\bibfield
  {journal} {\bibinfo  {journal} {Classical and Quantum Gravity}\ }\textbf
  {\bibinfo {volume} {29}},\ \bibinfo {pages} {184004} (\bibinfo {year}
  {2012})}\BibitemShut {NoStop}%
\bibitem [{\citenamefont {Fray}\ \emph {et~al.}(2004)\citenamefont {Fray},
  \citenamefont {Diez}, \citenamefont {H{\"a}nsch},\ and\ \citenamefont
  {Weitz}}]{Haensch:2004}%
  \BibitemOpen
  \bibfield  {author} {\bibinfo {author} {\bibfnamefont {S.}~\bibnamefont
  {Fray}}, \bibinfo {author} {\bibfnamefont {C.~A.}\ \bibnamefont {Diez}},
  \bibinfo {author} {\bibfnamefont {T.~W.}\ \bibnamefont {H{\"a}nsch}}, \ and\
  \bibinfo {author} {\bibfnamefont {M.}~\bibnamefont {Weitz}},\ }\href
  {\doibase 10.1103/PhysRevLett.93.240404} {\bibfield  {journal} {\bibinfo
  {journal} {Physical Review Letters}\ }\textbf {\bibinfo {volume} {93}},\
  \bibinfo {pages} {240404} (\bibinfo {year} {2004})}\BibitemShut {NoStop}%
\bibitem [{\citenamefont {Schlippert}\ \emph
  {et~al.}(2014{\natexlab{a}})\citenamefont {Schlippert}, \citenamefont
  {Hartwig}, \citenamefont {Albers}, \citenamefont {Richardson}, \citenamefont
  {Schubert}, \citenamefont {Roura}, \citenamefont {Schleich}, \citenamefont
  {Ertmer},\ and\ \citenamefont {Rasel}}]{Schlippert:2014}%
  \BibitemOpen
  \bibfield  {author} {\bibinfo {author} {\bibfnamefont {D.}~\bibnamefont
  {Schlippert}}, \bibinfo {author} {\bibfnamefont {J.}~\bibnamefont {Hartwig}},
  \bibinfo {author} {\bibfnamefont {H.}~\bibnamefont {Albers}}, \bibinfo
  {author} {\bibfnamefont {L.}~\bibnamefont {Richardson}}, \bibinfo {author}
  {\bibfnamefont {C.}~\bibnamefont {Schubert}}, \bibinfo {author}
  {\bibfnamefont {A.}~\bibnamefont {Roura}}, \bibinfo {author} {\bibfnamefont
  {W.}~\bibnamefont {Schleich}}, \bibinfo {author} {\bibfnamefont
  {W.}~\bibnamefont {Ertmer}}, \ and\ \bibinfo {author} {\bibfnamefont
  {E.}~\bibnamefont {Rasel}},\ }\href {\doibase 10.1103/PhysRevLett.112.203002}
  {\bibfield  {journal} {\bibinfo  {journal} {Physical Review Letters}\
  }\textbf {\bibinfo {volume} {112}},\ \bibinfo {pages} {203002} (\bibinfo
  {year} {2014}{\natexlab{a}})}\BibitemShut {NoStop}%
\bibitem [{\citenamefont {Zhou}\ \emph {et~al.}(2015)\citenamefont {Zhou},
  \citenamefont {Long}, \citenamefont {Tang}, \citenamefont {Chen},
  \citenamefont {Gao}, \citenamefont {Peng}, \citenamefont {Duan},
  \citenamefont {Zhong}, \citenamefont {Xiong}, \citenamefont {Wang} \emph
  {et~al.}}]{zhou2015test}%
  \BibitemOpen
  \bibfield  {author} {\bibinfo {author} {\bibfnamefont {L.}~\bibnamefont
  {Zhou}}, \bibinfo {author} {\bibfnamefont {S.}~\bibnamefont {Long}}, \bibinfo
  {author} {\bibfnamefont {B.}~\bibnamefont {Tang}}, \bibinfo {author}
  {\bibfnamefont {X.}~\bibnamefont {Chen}}, \bibinfo {author} {\bibfnamefont
  {F.}~\bibnamefont {Gao}}, \bibinfo {author} {\bibfnamefont {W.}~\bibnamefont
  {Peng}}, \bibinfo {author} {\bibfnamefont {W.}~\bibnamefont {Duan}}, \bibinfo
  {author} {\bibfnamefont {J.}~\bibnamefont {Zhong}}, \bibinfo {author}
  {\bibfnamefont {Z.}~\bibnamefont {Xiong}}, \bibinfo {author} {\bibfnamefont
  {J.}~\bibnamefont {Wang}},  \emph {et~al.},\ }\href {\doibase
  10.1103/PhysRevLett.115.013004} {\bibfield  {journal} {\bibinfo  {journal}
  {Physical Review Letters}\ }\textbf {\bibinfo {volume} {115}},\ \bibinfo
  {pages} {013004} (\bibinfo {year} {2015})}\BibitemShut {NoStop}%
\bibitem [{\citenamefont {Touboul}\ \emph {et~al.}(2017)\citenamefont
  {Touboul}, \citenamefont {M{\'e}tris}, \citenamefont {Rodrigues},
  \citenamefont {Andr{\'e}}, \citenamefont {Baghi}, \citenamefont {Berg{\'e}},
  \citenamefont {Boulanger}, \citenamefont {Bremer}, \citenamefont {Carle},
  \citenamefont {Chhun} \emph {et~al.}}]{touboul2017microscope}%
  \BibitemOpen
  \bibfield  {author} {\bibinfo {author} {\bibfnamefont {P.}~\bibnamefont
  {Touboul}}, \bibinfo {author} {\bibfnamefont {G.}~\bibnamefont {M{\'e}tris}},
  \bibinfo {author} {\bibfnamefont {M.}~\bibnamefont {Rodrigues}}, \bibinfo
  {author} {\bibfnamefont {Y.}~\bibnamefont {Andr{\'e}}}, \bibinfo {author}
  {\bibfnamefont {Q.}~\bibnamefont {Baghi}}, \bibinfo {author} {\bibfnamefont
  {J.}~\bibnamefont {Berg{\'e}}}, \bibinfo {author} {\bibfnamefont
  {D.}~\bibnamefont {Boulanger}}, \bibinfo {author} {\bibfnamefont
  {S.}~\bibnamefont {Bremer}}, \bibinfo {author} {\bibfnamefont
  {P.}~\bibnamefont {Carle}}, \bibinfo {author} {\bibfnamefont
  {R.}~\bibnamefont {Chhun}},  \emph {et~al.},\ }\href {\doibase
  10.1103/PhysRevLett.119.231101} {\bibfield  {journal} {\bibinfo  {journal}
  {Physical review letters}\ }\textbf {\bibinfo {volume} {119}},\ \bibinfo
  {pages} {231101} (\bibinfo {year} {2017})}\BibitemShut {NoStop}%
\bibitem [{\citenamefont {Archibald}\ \emph {et~al.}(2018)\citenamefont
  {Archibald}, \citenamefont {Gusinskaia}, \citenamefont {Hessels},
  \citenamefont {Deller}, \citenamefont {Kaplan}, \citenamefont {Lorimer},
  \citenamefont {Lynch}, \citenamefont {Ransom},\ and\ \citenamefont
  {Stairs}}]{archibald2018universality}%
  \BibitemOpen
  \bibfield  {author} {\bibinfo {author} {\bibfnamefont {A.~M.}\ \bibnamefont
  {Archibald}}, \bibinfo {author} {\bibfnamefont {N.~V.}\ \bibnamefont
  {Gusinskaia}}, \bibinfo {author} {\bibfnamefont {J.~W.}\ \bibnamefont
  {Hessels}}, \bibinfo {author} {\bibfnamefont {A.~T.}\ \bibnamefont {Deller}},
  \bibinfo {author} {\bibfnamefont {D.~L.}\ \bibnamefont {Kaplan}}, \bibinfo
  {author} {\bibfnamefont {D.~R.}\ \bibnamefont {Lorimer}}, \bibinfo {author}
  {\bibfnamefont {R.~S.}\ \bibnamefont {Lynch}}, \bibinfo {author}
  {\bibfnamefont {S.~M.}\ \bibnamefont {Ransom}}, \ and\ \bibinfo {author}
  {\bibfnamefont {I.~H.}\ \bibnamefont {Stairs}},\ }\href {\doibase
  10.1038/s41586-018-0265-1} {\bibfield  {journal} {\bibinfo  {journal}
  {Nature}\ }\textbf {\bibinfo {volume} {559}},\ \bibinfo {pages} {73}
  (\bibinfo {year} {2018})}\BibitemShut {NoStop}%
\bibitem [{\citenamefont {Zych}\ \emph {et~al.}(2011)\citenamefont {Zych},
  \citenamefont {Costa}, \citenamefont {Pikovski},\ and\ \citenamefont
  {Brukner}}]{Zych:2011}%
  \BibitemOpen
  \bibfield  {author} {\bibinfo {author} {\bibfnamefont {M.}~\bibnamefont
  {Zych}}, \bibinfo {author} {\bibfnamefont {F.}~\bibnamefont {Costa}},
  \bibinfo {author} {\bibfnamefont {I.}~\bibnamefont {Pikovski}}, \ and\
  \bibinfo {author} {\bibfnamefont {{\v{C}}.}~\bibnamefont {Brukner}},\ }\href
  {\doibase 10.1038/ncomms1498} {\bibfield  {journal} {\bibinfo  {journal}
  {Nature Communications}\ }\textbf {\bibinfo {volume} {2}},\ \bibinfo {pages}
  {505} (\bibinfo {year} {2011})}\BibitemShut {NoStop}%
\bibitem [{\citenamefont {Zych}\ \emph {et~al.}(2012)\citenamefont {Zych},
  \citenamefont {Costa}, \citenamefont {Pikovski}, \citenamefont {Ralph},\ and\
  \citenamefont {Brukner}}]{Zych:2012}%
  \BibitemOpen
  \bibfield  {author} {\bibinfo {author} {\bibfnamefont {M.}~\bibnamefont
  {Zych}}, \bibinfo {author} {\bibfnamefont {F.}~\bibnamefont {Costa}},
  \bibinfo {author} {\bibfnamefont {I.}~\bibnamefont {Pikovski}}, \bibinfo
  {author} {\bibfnamefont {T.~C.}\ \bibnamefont {Ralph}}, \ and\ \bibinfo
  {author} {\bibfnamefont {{\v C}.}~\bibnamefont {Brukner}},\ }\href {\doibase
  10.1088/0264-9381/29/22/224010} {\bibfield  {journal} {\bibinfo  {journal}
  {Classical and Quantum Gravity}\ }\textbf {\bibinfo {volume} {29}},\ \bibinfo
  {pages} {224010} (\bibinfo {year} {2012})}\BibitemShut {NoStop}%
\bibitem [{\citenamefont {Zych}\ \emph {et~al.}(2016)\citenamefont {Zych},
  \citenamefont {Pikovski}, \citenamefont {Costa},\ and\ \citenamefont
  {Brukner}}]{Zych2016}%
  \BibitemOpen
  \bibfield  {author} {\bibinfo {author} {\bibfnamefont {M.}~\bibnamefont
  {Zych}}, \bibinfo {author} {\bibfnamefont {I.}~\bibnamefont {Pikovski}},
  \bibinfo {author} {\bibfnamefont {F.}~\bibnamefont {Costa}}, \ and\ \bibinfo
  {author} {\bibfnamefont {{\v C}.}~\bibnamefont {Brukner}},\ }\href {\doibase
  10.1088/1742-6596/723/1/012044} {\bibfield  {journal} {\bibinfo  {journal}
  {Journal of Physics: Conference Series}\ }\textbf {\bibinfo {volume} {723}},\
  \bibinfo {pages} {012044} (\bibinfo {year} {2016})}\BibitemShut {NoStop}%
\bibitem [{\citenamefont {Margalit}\ \emph {et~al.}(2015)\citenamefont
  {Margalit}, \citenamefont {Zhou}, \citenamefont {Machluf}, \citenamefont
  {Rohrlich}, \citenamefont {Japha},\ and\ \citenamefont
  {Folman}}]{margalit2015self}%
  \BibitemOpen
  \bibfield  {author} {\bibinfo {author} {\bibfnamefont {Y.}~\bibnamefont
  {Margalit}}, \bibinfo {author} {\bibfnamefont {Z.}~\bibnamefont {Zhou}},
  \bibinfo {author} {\bibfnamefont {S.}~\bibnamefont {Machluf}}, \bibinfo
  {author} {\bibfnamefont {D.}~\bibnamefont {Rohrlich}}, \bibinfo {author}
  {\bibfnamefont {Y.}~\bibnamefont {Japha}}, \ and\ \bibinfo {author}
  {\bibfnamefont {R.}~\bibnamefont {Folman}},\ }\href {\doibase
  10.1126/science.aac6498} {\bibfield  {journal} {\bibinfo  {journal}
  {Science}\ }\textbf {\bibinfo {volume} {349}},\ \bibinfo {pages} {1205}
  (\bibinfo {year} {2015})}\BibitemShut {NoStop}%
\bibitem [{\citenamefont {Bushev}\ \emph {et~al.}(2016)\citenamefont {Bushev},
  \citenamefont {Cole}, \citenamefont {Sholokhov}, \citenamefont {Kukharchyk},\
  and\ \citenamefont {Zych}}]{bushev2016single}%
  \BibitemOpen
  \bibfield  {author} {\bibinfo {author} {\bibfnamefont {P.~A.}\ \bibnamefont
  {Bushev}}, \bibinfo {author} {\bibfnamefont {J.~H.}\ \bibnamefont {Cole}},
  \bibinfo {author} {\bibfnamefont {D.}~\bibnamefont {Sholokhov}}, \bibinfo
  {author} {\bibfnamefont {N.}~\bibnamefont {Kukharchyk}}, \ and\ \bibinfo
  {author} {\bibfnamefont {M.}~\bibnamefont {Zych}},\ }\href {\doibase
  10.1088/1367-2630/18/9/093050} {\bibfield  {journal} {\bibinfo  {journal}
  {New Journal of Physics}\ }\textbf {\bibinfo {volume} {18}},\ \bibinfo
  {pages} {093050} (\bibinfo {year} {2016})}\BibitemShut {NoStop}%
\bibitem [{\citenamefont {Pikovski}\ \emph {et~al.}(2015)\citenamefont
  {Pikovski}, \citenamefont {Zych}, \citenamefont {Costa},\ and\ \citenamefont
  {Brukner}}]{pikovskiuniversal2015}%
  \BibitemOpen
  \bibfield  {author} {\bibinfo {author} {\bibfnamefont {I.}~\bibnamefont
  {Pikovski}}, \bibinfo {author} {\bibfnamefont {M.}~\bibnamefont {Zych}},
  \bibinfo {author} {\bibfnamefont {F.}~\bibnamefont {Costa}}, \ and\ \bibinfo
  {author} {\bibfnamefont {{\v C}.}~\bibnamefont {Brukner}},\ }\href {\doibase
  10.1038/nphys3366} {\bibfield  {journal} {\bibinfo  {journal} {Nature
  Physics}\ }\textbf {\bibinfo {volume} {11}},\ \bibinfo {pages} {668}
  (\bibinfo {year} {2015})}\BibitemShut {NoStop}%
\bibitem [{\citenamefont {Adler}\ and\ \citenamefont {Bassi}(2016)}]{ref:AB}%
  \BibitemOpen
  \bibfield  {author} {\bibinfo {author} {\bibfnamefont {S.~L.}\ \bibnamefont
  {Adler}}\ and\ \bibinfo {author} {\bibfnamefont {A.}~\bibnamefont {Bassi}},\
  }\href {\doibase http://dx.doi.org/10.1016/j.physleta.2015.10.064} {\bibfield
   {journal} {\bibinfo  {journal} {Physics Letters A}\ }\textbf {\bibinfo
  {volume} {380}},\ \bibinfo {pages} {390 } (\bibinfo {year}
  {2016})}\BibitemShut {NoStop}%
\bibitem [{\citenamefont {Pikovski}\ \emph {et~al.}(2017)\citenamefont
  {Pikovski}, \citenamefont {Zych}, \citenamefont {Costa},\ and\ \citenamefont
  {Brukner}}]{PikovskiTime2017}%
  \BibitemOpen
  \bibfield  {author} {\bibinfo {author} {\bibfnamefont {I.}~\bibnamefont
  {Pikovski}}, \bibinfo {author} {\bibfnamefont {M.}~\bibnamefont {Zych}},
  \bibinfo {author} {\bibfnamefont {F.}~\bibnamefont {Costa}}, \ and\ \bibinfo
  {author} {\bibfnamefont {{\v C}.}~\bibnamefont {Brukner}},\ }\href {\doibase
  10.1088/1367-2630/aa5d92} {\bibfield  {journal} {\bibinfo  {journal} {New
  Journal of Physics}\ }\textbf {\bibinfo {volume} {19}},\ \bibinfo {pages}
  {025011} (\bibinfo {year} {2017})}\BibitemShut {NoStop}%
\bibitem [{\citenamefont {Korbicz}\ and\ \citenamefont
  {Tuziemski}(2017)}]{korbicz2017information}%
  \BibitemOpen
  \bibfield  {author} {\bibinfo {author} {\bibfnamefont {J.}~\bibnamefont
  {Korbicz}}\ and\ \bibinfo {author} {\bibfnamefont {J.}~\bibnamefont
  {Tuziemski}},\ }\href {\doibase 10.1007/s10714-017-2319-3} {\bibfield
  {journal} {\bibinfo  {journal} {General Relativity and Gravitation}\ }\textbf
  {\bibinfo {volume} {49}},\ \bibinfo {pages} {152} (\bibinfo {year}
  {2017})}\BibitemShut {NoStop}%
\bibitem [{\citenamefont {Zych}\ and\ \citenamefont
  {Brukner}(2018)}]{zych2015quantum}%
  \BibitemOpen
  \bibfield  {author} {\bibinfo {author} {\bibfnamefont {M.}~\bibnamefont
  {Zych}}\ and\ \bibinfo {author} {\bibfnamefont {{\v C}.}~\bibnamefont
  {Brukner}},\ }\href {\doibase 10.1038/s41567-018-0197-6} {\bibfield
  {journal} {\bibinfo  {journal} {Nature Physics}\ } (\bibinfo {year} {2018}),\
  10.1038/s41567-018-0197-6}\BibitemShut {NoStop}%
\bibitem [{\citenamefont {Zych}(2017)}]{zych2015PhD}%
  \BibitemOpen
  \bibfield  {author} {\bibinfo {author} {\bibfnamefont {M.}~\bibnamefont
  {Zych}},\ }\href {\doibase 10.1007/978-3-319-53192-2} {\emph {\bibinfo
  {title} {Quantum systems under gravitational time dilation}}},\ Springer
  Theses\ (\bibinfo  {publisher} {Springer},\ \bibinfo {year}
  {2017})\BibitemShut {NoStop}%
\bibitem [{\citenamefont {Rosi}\ \emph {et~al.}(2017)\citenamefont {Rosi},
  \citenamefont {D'Amico}, \citenamefont {Cacciapuoti}, \citenamefont
  {Sorrentino}, \citenamefont {Prevedelli}, \citenamefont {Zych}, \citenamefont
  {Brukner},\ and\ \citenamefont {Tino}}]{rosi2017quantum}%
  \BibitemOpen
  \bibfield  {author} {\bibinfo {author} {\bibfnamefont {G.}~\bibnamefont
  {Rosi}}, \bibinfo {author} {\bibfnamefont {G.}~\bibnamefont {D'Amico}},
  \bibinfo {author} {\bibfnamefont {L.}~\bibnamefont {Cacciapuoti}}, \bibinfo
  {author} {\bibfnamefont {F.}~\bibnamefont {Sorrentino}}, \bibinfo {author}
  {\bibfnamefont {M.}~\bibnamefont {Prevedelli}}, \bibinfo {author}
  {\bibfnamefont {M.}~\bibnamefont {Zych}}, \bibinfo {author} {\bibfnamefont
  {{\v{C}}.}~\bibnamefont {Brukner}}, \ and\ \bibinfo {author} {\bibfnamefont
  {G.}~\bibnamefont {Tino}},\ }\href {\doibase 10.1038/ncomms15529} {\bibfield
  {journal} {\bibinfo  {journal} {Nature communications}\ }\textbf {\bibinfo
  {volume} {8}},\ \bibinfo {pages} {15529} (\bibinfo {year}
  {2017})}\BibitemShut {NoStop}%
\bibitem [{\citenamefont {{Orlando}}\ \emph {et~al.}(2016)\citenamefont
  {{Orlando}}, \citenamefont {{Mann}}, \citenamefont {{Modi}},\ and\
  \citenamefont {{Pollock}}}]{2016CQGra..33sLT01O}%
  \BibitemOpen
  \bibfield  {author} {\bibinfo {author} {\bibfnamefont {P.~J.}\ \bibnamefont
  {{Orlando}}}, \bibinfo {author} {\bibfnamefont {R.~B.}\ \bibnamefont
  {{Mann}}}, \bibinfo {author} {\bibfnamefont {K.}~\bibnamefont {{Modi}}}, \
  and\ \bibinfo {author} {\bibfnamefont {F.~A.}\ \bibnamefont {{Pollock}}},\
  }\href {\doibase 10.1088/0264-9381/33/19/19LT01} {\bibfield  {journal}
  {\bibinfo  {journal} {Classical and Quantum Gravity}\ }\textbf {\bibinfo
  {volume} {33}},\ \bibinfo {eid} {19LT01} (\bibinfo {year}
  {2016})}\BibitemShut {NoStop}%
\bibitem [{\citenamefont {Geiger}\ and\ \citenamefont
  {Trupke}(2018)}]{geiger2018proposal}%
  \BibitemOpen
  \bibfield  {author} {\bibinfo {author} {\bibfnamefont {R.}~\bibnamefont
  {Geiger}}\ and\ \bibinfo {author} {\bibfnamefont {M.}~\bibnamefont
  {Trupke}},\ }\href {\doibase 10.1103/PhysRevLett.120.043602} {\bibfield
  {journal} {\bibinfo  {journal} {Physical Review Letters}\ }\textbf {\bibinfo
  {volume} {120}},\ \bibinfo {pages} {043602} (\bibinfo {year}
  {2018})}\BibitemShut {NoStop}%
\bibitem [{\citenamefont {Eddington}\ and\ \citenamefont
  {Clark}(1938)}]{eddington1938problem}%
  \BibitemOpen
  \bibfield  {author} {\bibinfo {author} {\bibfnamefont {A.~S.}\ \bibnamefont
  {Eddington}}\ and\ \bibinfo {author} {\bibfnamefont {G.~L.}\ \bibnamefont
  {Clark}},\ }\href {\doibase 10.1098/rspa.1938.0104} {\bibfield  {journal}
  {\bibinfo  {journal} {Proc. R. Soc. Lond. A}\ }\textbf {\bibinfo {volume}
  {166}},\ \bibinfo {pages} {465} (\bibinfo {year} {1938})}\BibitemShut
  {NoStop}%
\bibitem [{\citenamefont {Nordtvedt}(1970)}]{nordtvedt1970gravitational}%
  \BibitemOpen
  \bibfield  {author} {\bibinfo {author} {\bibfnamefont {K.}~\bibnamefont
  {Nordtvedt}},\ }\href {\doibase 10.1007/BF02412754} {\bibfield  {journal}
  {\bibinfo  {journal} {International Journal of Theoretical Physics}\ }\textbf
  {\bibinfo {volume} {3}},\ \bibinfo {pages} {133} (\bibinfo {year}
  {1970})}\BibitemShut {NoStop}%
\bibitem [{\citenamefont {Lightman}\ and\ \citenamefont
  {Lee}(1973)}]{lightman1973restricted}%
  \BibitemOpen
  \bibfield  {author} {\bibinfo {author} {\bibfnamefont {A.~P.}\ \bibnamefont
  {Lightman}}\ and\ \bibinfo {author} {\bibfnamefont {D.~L.}\ \bibnamefont
  {Lee}},\ }\href {\doibase 10.1103/PhysRevD.8.364} {\bibfield  {journal}
  {\bibinfo  {journal} {Physical Review D}\ }\textbf {\bibinfo {volume} {8}},\
  \bibinfo {pages} {364} (\bibinfo {year} {1973})}\BibitemShut {NoStop}%
\bibitem [{\citenamefont {Nordtvedt}(1974)}]{nordtvedt1974equation}%
  \BibitemOpen
  \bibfield  {author} {\bibinfo {author} {\bibfnamefont {K.}~\bibnamefont
  {Nordtvedt}},\ }\href {\doibase 10.1007/BF01810699} {\bibfield  {journal}
  {\bibinfo  {journal} {International Journal of Theoretical Physics}\ }\textbf
  {\bibinfo {volume} {9}},\ \bibinfo {pages} {269} (\bibinfo {year}
  {1974})}\BibitemShut {NoStop}%
\bibitem [{\citenamefont {Nordtvedt}(1994)}]{Nordtvedt:1994PostNewtonian}%
  \BibitemOpen
  \bibfield  {author} {\bibinfo {author} {\bibfnamefont {K.}~\bibnamefont
  {Nordtvedt}},\ }\href {\doibase https://doi.org/10.1088/0264-9381/11/6A/009}
  {\bibfield  {journal} {\bibinfo  {journal} {Classical and Quantum Gravity}\
  }\textbf {\bibinfo {volume} {11}},\ \bibinfo {pages} {A119} (\bibinfo {year}
  {1994})}\BibitemShut {NoStop}%
\bibitem [{\citenamefont {Carlip}(1998)}]{Carlip:1999KE}%
  \BibitemOpen
  \bibfield  {author} {\bibinfo {author} {\bibfnamefont {S.}~\bibnamefont
  {Carlip}},\ }\href {\doibase 10.1119/1.18885} {\bibfield  {journal} {\bibinfo
   {journal} {American Journal of Physics}\ }\textbf {\bibinfo {volume} {66}},\
  \bibinfo {pages} {409} (\bibinfo {year} {1998})}\BibitemShut {NoStop}%
\bibitem [{\citenamefont {Fischbach}\ \emph {et~al.}(1981)\citenamefont
  {Fischbach}, \citenamefont {Freeman},\ and\ \citenamefont
  {Cheng}}]{PhysRevD.23.2157}%
  \BibitemOpen
  \bibfield  {author} {\bibinfo {author} {\bibfnamefont {E.}~\bibnamefont
  {Fischbach}}, \bibinfo {author} {\bibfnamefont {B.~S.}\ \bibnamefont
  {Freeman}}, \ and\ \bibinfo {author} {\bibfnamefont {W.-K.}\ \bibnamefont
  {Cheng}},\ }\href {\doibase 10.1103/PhysRevD.23.2157} {\bibfield  {journal}
  {\bibinfo  {journal} {Physical Review D}\ }\textbf {\bibinfo {volume} {23}},\
  \bibinfo {pages} {2157} (\bibinfo {year} {1981})}\BibitemShut {NoStop}%
\bibitem [{\citenamefont {Lebed}(2013)}]{Lebed2013}%
  \BibitemOpen
  \bibfield  {author} {\bibinfo {author} {\bibfnamefont {A.~G.}\ \bibnamefont
  {Lebed}},\ }\href {\doibase 10.2478/s11534-013-0302-5} {\bibfield  {journal}
  {\bibinfo  {journal} {Central European Journal of Physics}\ }\textbf
  {\bibinfo {volume} {11}},\ \bibinfo {pages} {969} (\bibinfo {year}
  {2013})}\BibitemShut {NoStop}%
\bibitem [{\citenamefont {Arnowitt}\ \emph {et~al.}(2008)\citenamefont
  {Arnowitt}, \citenamefont {Deser},\ and\ \citenamefont {Misner}}]{ADM:2008}%
  \BibitemOpen
  \bibfield  {author} {\bibinfo {author} {\bibfnamefont {R.}~\bibnamefont
  {Arnowitt}}, \bibinfo {author} {\bibfnamefont {S.}~\bibnamefont {Deser}}, \
  and\ \bibinfo {author} {\bibfnamefont {C.~W.}\ \bibnamefont {Misner}},\
  }\href {\doibase 10.1007/s10714-008-0661-1} {\bibfield  {journal} {\bibinfo
  {journal} {General Relativity and Gravitation}\ }\textbf {\bibinfo {volume}
  {40}},\ \bibinfo {pages} {1997} (\bibinfo {year} {2008})}\BibitemShut
  {NoStop}%
\bibitem [{\citenamefont {Brill}\ and\ \citenamefont
  {Lindquist}(1963)}]{Brill:1963Energy}%
  \BibitemOpen
  \bibfield  {author} {\bibinfo {author} {\bibfnamefont {D.~R.}\ \bibnamefont
  {Brill}}\ and\ \bibinfo {author} {\bibfnamefont {R.~W.}\ \bibnamefont
  {Lindquist}},\ }\href {\doibase 10.1103/PhysRev.131.471} {\bibfield
  {journal} {\bibinfo  {journal} {Phys. Rev.}\ }\textbf {\bibinfo {volume}
  {131}},\ \bibinfo {pages} {471} (\bibinfo {year} {1963})}\BibitemShut
  {NoStop}%
\bibitem [{Note1()}]{Note1}%
  \BibitemOpen
  \bibinfo {note} {Provided that tidal forces are negligible -- the usual
  assumption under which the equivalence principle is required to
  hold.}\BibitemShut {Stop}%
\bibitem [{\citenamefont {Weinberg}(1972)}]{WeinbergGR}%
  \BibitemOpen
  \bibfield  {author} {\bibinfo {author} {\bibfnamefont {S.}~\bibnamefont
  {Weinberg}},\ }\href@noop {} {\emph {\bibinfo {title} {Gravitation and
  cosmology:}}}{ Principle and applications of general theory of relativity}\
  (\bibinfo  {publisher} {John Wiley and Sons, Inc., New York},\ \bibinfo
  {year} {1972})\BibitemShut {NoStop}%
\bibitem [{\citenamefont {Landau}\ and\ \citenamefont
  {Lifshitz}(1975)}]{Landau:1975}%
  \BibitemOpen
  \bibfield  {author} {\bibinfo {author} {\bibfnamefont {L.~D.}\ \bibnamefont
  {Landau}}\ and\ \bibinfo {author} {\bibfnamefont {E.~M.}\ \bibnamefont
  {Lifshitz}},\ }\href@noop {} {\emph {\bibinfo {title} {The Classical Theory
  of Fields}}}\ (\bibinfo  {publisher} {Pergamon Press Ltd.},\ \bibinfo
  {address} {Oxford},\ \bibinfo {year} {1975})\BibitemShut {NoStop}%
\bibitem [{Note2()}]{Note2}%
  \BibitemOpen
  \bibinfo {note} {The parameter $t$ is operationally defined as the time
  measured by a clock at rest in the origin of the reference frame with respect
  to which the spatial coordinates of the particles are defined.}\BibitemShut
  {Stop}%
\bibitem [{\citenamefont {{Damour}}(1987)}]{damour1987problem}%
  \BibitemOpen
  \bibfield  {author} {\bibinfo {author} {\bibfnamefont {T.}~\bibnamefont
  {{Damour}}},\ }\enquote {\bibinfo {title} {{The problem of motion in
  Newtonian and Einsteinian gravity.}}}\ in\ \href@noop {} {\emph {\bibinfo
  {booktitle} {Three Hundred Years of Gravitation}}},\ \bibinfo {editor}
  {edited by\ \bibinfo {editor} {\bibfnamefont {S.~W.}\ \bibnamefont
  {{Hawking}}}\ and\ \bibinfo {editor} {\bibfnamefont {W.}~\bibnamefont
  {{Israel}}}}\ (\bibinfo {year} {1987})\ pp.\ \bibinfo {pages}
  {128--198}\BibitemShut {NoStop}%
\bibitem [{\citenamefont {Brumberg}\ and\ \citenamefont
  {Kopejkin}(1989)}]{Brumberg1989}%
  \BibitemOpen
  \bibfield  {author} {\bibinfo {author} {\bibfnamefont {V.~A.}\ \bibnamefont
  {Brumberg}}\ and\ \bibinfo {author} {\bibfnamefont {S.~M.}\ \bibnamefont
  {Kopejkin}},\ }\href {\doibase 10.1007/BF02888894} {\bibfield  {journal}
  {\bibinfo  {journal} {Il Nuovo Cimento B (1971-1996)}\ }\textbf {\bibinfo
  {volume} {103}},\ \bibinfo {pages} {63} (\bibinfo {year} {1989})}\BibitemShut
  {NoStop}%
\bibitem [{\citenamefont {Damour}\ \emph {et~al.}(1991)\citenamefont {Damour},
  \citenamefont {Soffel},\ and\ \citenamefont {Xu}}]{PhysRevD.43.3273}%
  \BibitemOpen
  \bibfield  {author} {\bibinfo {author} {\bibfnamefont {T.}~\bibnamefont
  {Damour}}, \bibinfo {author} {\bibfnamefont {M.}~\bibnamefont {Soffel}}, \
  and\ \bibinfo {author} {\bibfnamefont {C.}~\bibnamefont {Xu}},\ }\href
  {\doibase 10.1103/PhysRevD.43.3273} {\bibfield  {journal} {\bibinfo
  {journal} {Phys. Rev. D}\ }\textbf {\bibinfo {volume} {43}},\ \bibinfo
  {pages} {3273} (\bibinfo {year} {1991})}\BibitemShut {NoStop}%
\bibitem [{\citenamefont {Damour}\ \emph {et~al.}(1992)\citenamefont {Damour},
  \citenamefont {Soffel},\ and\ \citenamefont {Xu}}]{PhysRevD.45.1017}%
  \BibitemOpen
  \bibfield  {author} {\bibinfo {author} {\bibfnamefont {T.}~\bibnamefont
  {Damour}}, \bibinfo {author} {\bibfnamefont {M.}~\bibnamefont {Soffel}}, \
  and\ \bibinfo {author} {\bibfnamefont {C.}~\bibnamefont {Xu}},\ }\href
  {\doibase 10.1103/PhysRevD.45.1017} {\bibfield  {journal} {\bibinfo
  {journal} {Phys. Rev. D}\ }\textbf {\bibinfo {volume} {45}},\ \bibinfo
  {pages} {1017} (\bibinfo {year} {1992})}\BibitemShut {NoStop}%
\bibitem [{\citenamefont {Damour}\ \emph {et~al.}(1994)\citenamefont {Damour},
  \citenamefont {Soffel},\ and\ \citenamefont {Xu}}]{PhysRevD.49.618}%
  \BibitemOpen
  \bibfield  {author} {\bibinfo {author} {\bibfnamefont {T.}~\bibnamefont
  {Damour}}, \bibinfo {author} {\bibfnamefont {M.}~\bibnamefont {Soffel}}, \
  and\ \bibinfo {author} {\bibfnamefont {C.}~\bibnamefont {Xu}},\ }\href
  {\doibase 10.1103/PhysRevD.49.618} {\bibfield  {journal} {\bibinfo  {journal}
  {Phys. Rev. D}\ }\textbf {\bibinfo {volume} {49}},\ \bibinfo {pages} {618}
  (\bibinfo {year} {1994})}\BibitemShut {NoStop}%
\bibitem [{\citenamefont {Dixon}(1964)}]{Dixon:1964}%
  \BibitemOpen
  \bibfield  {author} {\bibinfo {author} {\bibfnamefont {W.}~\bibnamefont
  {Dixon}},\ }\href {\doibase https://doi.org/10.1007/BF02734579} {\bibfield
  {journal} {\bibinfo  {journal} {Il Nuovo Cimento}\ }\textbf {\bibinfo
  {volume} {34}},\ \bibinfo {pages} {317} (\bibinfo {year} {1964})}\BibitemShut
  {NoStop}%
\bibitem [{\citenamefont {Beiglb{\"o}ck}(1967)}]{beiglbock1967}%
  \BibitemOpen
  \bibfield  {author} {\bibinfo {author} {\bibfnamefont {W.}~\bibnamefont
  {Beiglb{\"o}ck}},\ }\href {\doibase 10.1007/BF01646841} {\bibfield  {journal}
  {\bibinfo  {journal} {Comm. Math. Phys.}\ }\textbf {\bibinfo {volume} {5}},\
  \bibinfo {pages} {106} (\bibinfo {year} {1967})}\BibitemShut {NoStop}%
\bibitem [{\citenamefont {Einstein}\ \emph {et~al.}(1938)\citenamefont
  {Einstein}, \citenamefont {Infeld},\ and\ \citenamefont
  {Hoffmann}}]{einstein1938gravitational}%
  \BibitemOpen
  \bibfield  {author} {\bibinfo {author} {\bibfnamefont {A.}~\bibnamefont
  {Einstein}}, \bibinfo {author} {\bibfnamefont {L.}~\bibnamefont {Infeld}}, \
  and\ \bibinfo {author} {\bibfnamefont {B.}~\bibnamefont {Hoffmann}},\ }\href
  {\doibase 10.2307/1968714} {\bibfield  {journal} {\bibinfo  {journal} {Annals
  of Mathematics}\ ,\ \bibinfo {pages} {65}} (\bibinfo {year}
  {1938})}\BibitemShut {NoStop}%
\bibitem [{Note3()}]{Note3}%
  \BibitemOpen
  \bibinfo {note} {To this order of approximation, the coordinates for the
  relativistic center of inertia are well-defined}\BibitemShut {NoStop}%
\bibitem [{\citenamefont {Scully}\ \emph {et~al.}(2018)\citenamefont {Scully},
  \citenamefont {Fulling}, \citenamefont {Lee}, \citenamefont {Page},
  \citenamefont {Schleich},\ and\ \citenamefont
  {Svidzinsky}}]{scully2018quantum}%
  \BibitemOpen
  \bibfield  {author} {\bibinfo {author} {\bibfnamefont {M.~O.}\ \bibnamefont
  {Scully}}, \bibinfo {author} {\bibfnamefont {S.}~\bibnamefont {Fulling}},
  \bibinfo {author} {\bibfnamefont {D.~M.}\ \bibnamefont {Lee}}, \bibinfo
  {author} {\bibfnamefont {D.~N.}\ \bibnamefont {Page}}, \bibinfo {author}
  {\bibfnamefont {W.~P.}\ \bibnamefont {Schleich}}, \ and\ \bibinfo {author}
  {\bibfnamefont {A.~A.}\ \bibnamefont {Svidzinsky}},\ }\href {\doibase
  10.1073/pnas.1807703115} {\bibfield  {journal} {\bibinfo  {journal}
  {Proceedings of the National Academy of Sciences}\ }\textbf {\bibinfo
  {volume} {115}},\ \bibinfo {pages} {8131} (\bibinfo {year}
  {2018})}\BibitemShut {NoStop}%
\bibitem [{\citenamefont {Schlippert}\ \emph
  {et~al.}(2014{\natexlab{b}})\citenamefont {Schlippert}, \citenamefont
  {Hartwig}, \citenamefont {Albers}, \citenamefont {Richardson}, \citenamefont
  {Schubert}, \citenamefont {Roura}, \citenamefont {Schleich}, \citenamefont
  {Ertmer},\ and\ \citenamefont {Rasel}}]{schlippert2014quantum}%
  \BibitemOpen
  \bibfield  {author} {\bibinfo {author} {\bibfnamefont {D.}~\bibnamefont
  {Schlippert}}, \bibinfo {author} {\bibfnamefont {J.}~\bibnamefont {Hartwig}},
  \bibinfo {author} {\bibfnamefont {H.}~\bibnamefont {Albers}}, \bibinfo
  {author} {\bibfnamefont {L.~L.}\ \bibnamefont {Richardson}}, \bibinfo
  {author} {\bibfnamefont {C.}~\bibnamefont {Schubert}}, \bibinfo {author}
  {\bibfnamefont {A.}~\bibnamefont {Roura}}, \bibinfo {author} {\bibfnamefont
  {W.~P.}\ \bibnamefont {Schleich}}, \bibinfo {author} {\bibfnamefont
  {W.}~\bibnamefont {Ertmer}}, \ and\ \bibinfo {author} {\bibfnamefont {E.~M.}\
  \bibnamefont {Rasel}},\ }\href {\doibase 10.1103/PhysRevLett.112.203002}
  {\bibfield  {journal} {\bibinfo  {journal} {Physical Review Letters}\
  }\textbf {\bibinfo {volume} {112}},\ \bibinfo {pages} {203002} (\bibinfo
  {year} {2014}{\natexlab{b}})}\BibitemShut {NoStop}%
\bibitem [{\citenamefont {Sonnleitner}\ \emph {et~al.}(2017)\citenamefont
  {Sonnleitner}, \citenamefont {Trautmann},\ and\ \citenamefont
  {Barnett}}]{sonnleitner2017will}%
  \BibitemOpen
  \bibfield  {author} {\bibinfo {author} {\bibfnamefont {M.}~\bibnamefont
  {Sonnleitner}}, \bibinfo {author} {\bibfnamefont {N.}~\bibnamefont
  {Trautmann}}, \ and\ \bibinfo {author} {\bibfnamefont {S.~M.}\ \bibnamefont
  {Barnett}},\ }\href {\doibase 10.1103/PhysRevLett.118.053601} {\bibfield
  {journal} {\bibinfo  {journal} {Physical Review Letters}\ }\textbf {\bibinfo
  {volume} {118}},\ \bibinfo {pages} {053601} (\bibinfo {year}
  {2017})}\BibitemShut {NoStop}%
\bibitem [{\citenamefont {Sonnleitner}\ and\ \citenamefont
  {Barnett}(2018)}]{sonnleitner2018mass}%
  \BibitemOpen
  \bibfield  {author} {\bibinfo {author} {\bibfnamefont {M.}~\bibnamefont
  {Sonnleitner}}\ and\ \bibinfo {author} {\bibfnamefont {S.~M.}\ \bibnamefont
  {Barnett}},\ }\href {\doibase 10.1103/PhysRevA.98.042106} {\bibfield
  {journal} {\bibinfo  {journal} {Physical Review A}\ }\textbf {\bibinfo
  {volume} {98}},\ \bibinfo {pages} {042106} (\bibinfo {year}
  {2018})}\BibitemShut {NoStop}%
\end{thebibliography}

%

\end{document}